\documentclass[aps,prl,reprint,groupedaddress,twocolumn,showpacs]{revtex4-1}
\usepackage{graphicx}
\begin{document}
\title{Waves of DNA:  Propagating Excitations in Extended Nanoconfined Polymers}
\author{Alexander R. Klotz}
\affiliation{Department of Chemical Engineering, Massachusetts Institute of Technology}
\author{Hendrick W. de Haan}
\affiliation{Faculty of Science, University of Ontario Institute of Technology}
\author{Walter W. Reisner}
\affiliation{Department of Physics, McGill University}

\begin{abstract}
   We use a nanofluidic system to investigate the emergence of thermally driven collective phenomena along a single polymer chain.   In our approach, a single DNA molecule is confined in a nanofluidic slit etched with arrays of embedded nanocavities; the cavity lattice is designed so that a single chain occupies multiple cavities. Fluorescent video-microscopy data shows that waves of excess fluorescence propagate across the cavity-straddling molecule, corresponding to propagating fluctuations of contour overdensity in the cavities.  The waves are quantified by examining the correlation in intensity fluctuations between neighbouring cavities.  Correlations grow from an anti-correlated minimum to a correlated maximum before decaying, corresponding to a transfer of contour between neighbouring cavities at a fixed transfer time-scale. The observed dynamics can be modelled using Langevin dynamics simulations and a minimal lattice model of coupled diffusion.  This study shows how confinement-based sculpting of the polymer equilibrium configuration, by renormalizing the physical system into a series of discrete cavity states, can lead to new types of dynamic collective phenomena.
   
   \end{abstract}
\maketitle


\section{Introduction}

      When a polymer is confined in a structure with dimension below its equilibrium coil size (radius of gyration), the equilibrium polymer conformation will be altered by the enveloping device geometry \cite{walterreview}.  Simple nanochannel structures can be used to extend DNA for analysis for high-throughput genomic mapping \cite{reisner2005statics, hancao}.  There has been increasing interest, however, in the development of devices that feature more complex confinement, devices we term ``complex nanofluidic landscapes"  \cite{walterpnas, tetris}.  These structures, produced by incorporating embedded nanotopography in an open nanoslit, feature regions of locally varying dimensionality and confinement-scale, allowing for greater variety of physical phenomena and potential manipulations.  For example, nanogroove devices feature arrays of parallel grooves embedded in a slit; molecules can be driven in an extended conformation maintained perpendicular to the axis of applied force \cite{nanogroove1, nanogroove2}.  Nanocavity devices, featuring arrays of nanopits etched in a slit, have been investigated for their utility in molecular self-assembly \cite{walterpnas}, to control single polymer mobility \cite{delbon} and diffusion \cite{difres}, and as a probe of polymer confinement physics \cite{tetris}.
      
           Recently \cite{dimers} we showed that correlations in the fluctuations of contour in a molecule straddling two cavities can be understood in terms of harmonic modes arising from the confining potential sculpted by the nanofluidic geometry.  Making an analogy between statistical and quantum mechanics, the harmonic modes that give rise to local correlations correspond to bound states, whereas the propagating modes along the whole molecule correspond to scattering states.  Continuing with the harmonic oscillator analogy, extending the number of coupled oscillators to the large-$N$ limit gives rise to collective phenomena such as phonons.  We investigate phenomena analogous to phonons in our overdamped nanofluidic system by increasing the number of cavities occupied by the DNA.
  
  \begin{figure*}
	\centering
		\includegraphics[width=1\textwidth]{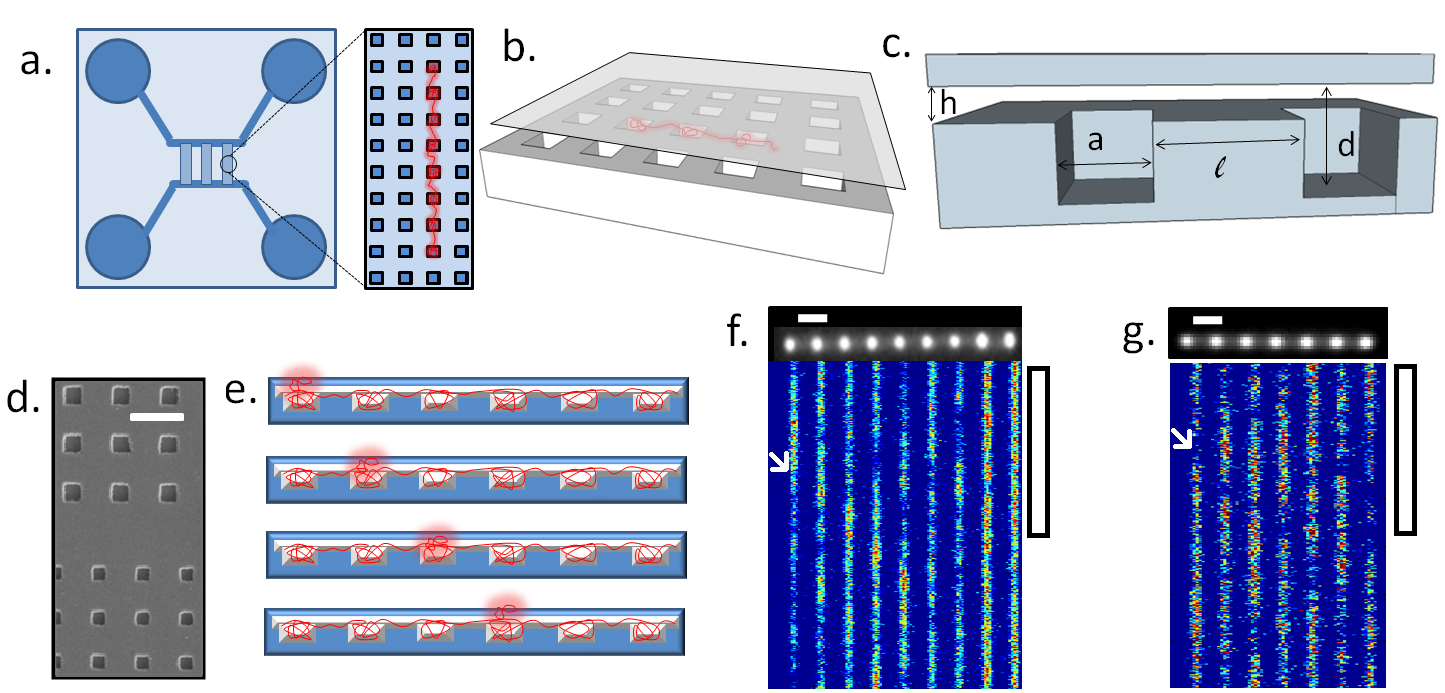}
	\label{fig:f1}
	\caption{Schematic of the experimental system. a. Diagram of the lab-on-a-chip device with the microfluidic reservoirs connected by nanofluidic slits. The slits are embedded with nanocavity lattices that act as entropic traps for the DNA. b. Oblique view of the nanoslit and nanocavity lattice. c. Side view of two cavities for definition of the geometric parameters of the system (typical values are $h=100 nm$, $d=200 nm$, $a=500 nm$ and $\ell=1000 nm$) d. Electron micrograph of two adjacent nanocavity arrays with 400\,nm and 300 \,nm cavity widths. Scale bar is one micron. e. Cartoon of a contour overdensity propagating along a molecule spanning seven pits. f. Fluorescence micrograph and time-space kymograph of a contour overdensity propagating through the molecule, manifesting itself as a brighter-than-average cavity. Horizontal scale bar is two microns, vertical scale bar is five seconds.  g. Fluorescence micrograph and time-space kymograph of a contour underdensity propagating through the molecule, manifesting itself as a dimmer-than-average cavity. Horizontal scale bar is two microns, vertical scale bar is five seconds.}
\end{figure*}    
      In this study genomic-length DNA molecules are confined to linear arrays of nanofluidic cavities embedded in a nanofluidic slit (see Fig. 1).  The device is constructed so that one molecule straddles and fills many cavities.  While the cavities are less confining than the surrounding slit, so it is entropically favourable for contour to accumulate in the cavities, excluded-volume interactions place an upper limit on the amount of contour that can be partitioned into a given cavity: a sufficiently long molecule can therefore fill multiple cavities.  In addition, self-exclusion confers a free energy cost to accumulating too much contour in a given cavity, leading to a self-exclusion driven relaxation process whereby contour is transferred to a neighbouring cavity via the connecting linker strand.  This cavity-to-cavity coupling leads to thermal excitations of molecular contour manifesting themselves as ``waves'' of excess fluorescent intensity propagating along the molecule, which can be seen in a video in the supplemental material (http://web.mit.edu/aklotz/www/SI1.gif).  This system provides a controlled manner to examine the emergence of collective phenomena, the observed density waves, from microscopic behaviour, the transfer of molecular contour between the cavities.  In these experiments, the network of nanofluidic entropic traps acts as a ``renormalizer'' to reduce the number of interacting components in the system to a small integer, reducing the infinite set of relaxation modes associated with Rouse-Zimm polymer dynamics into a finite set determined by the number of filled cavities.

 From an application point of view, emerging biotechnologies utilizing nanofluidic confinement to control or constrain DNA can benefit from a knowledge of the global behaviour of DNA contour redistribution. For example, when imaging a single region of a molecule, it is necessary to know for how long the image contains the same genetic information, to optimize the data acquisition parameters.  Such knowledge may guide the design of zero-mode-waveguide sequencing devices \cite{zeromode} or entropic sieves \cite{entropictrap}. Nanoslit molecular stretching devices are limited in observation time by the global transfer of contour across a slit between reservoirs \cite{koun}, and this time can be constrained using a knowledge of the internal contour transfer modes.   Future nanopore devices \cite{zhang2015fabrication} may rely on translocating DNA through multiple pores in serial to improve signal-to-noise, in which case it is necessary to understand how contour is transferred between the cavities on either side of a pore.

\section{Experimental Setup}

    The experimental system is a nanofluidic lab-on-a-chip device featuring a slit with a height on the order of 100\,nm that is embedded with a lattice of square cavities etched to twice the depth of the slit (see Fig. 1). The devices were fabricated through standard nanolithography and etching techniques (see \cite{walterpnas}).  Fluid is loaded into the nanofeatures through sandblasted holes that are interfaced to the nanoslit by microchannel arms (50$\mu$m wide, 1\,$\mu$m deep), positioned symmetrically on either side of the nanoslit.  The experiments were conducted with T4 DNA (166\.kbp) stained with YOYO-1 fluorescent dye at 10:1 base-pair:dye ratio.  In addition, we used the 43 kbp Charomid DNA (Wako), at the same dye loading, as an example of a molecule with circular topology.    The DNA is loaded in a 10 mM Tris buffer with 2\% beta-mercaptoethanol to suppress photobleaching and photo nicking.  Once loaded into the reservoirs,  pneumatic pressure is used to circulate DNA through the microchannel arms and then drive DNA from the microchannel into the nanoslit.   Once the molecules enter a region of the nanoslit containing nanocavity arrays, contour will be partitioned into adjacent cavities (see Fig. 1), forming self-assembled nanocavity conformations. Large DNA molecules such as T4 are liable to fragment due to shear or photo nicking, which we use to our advantage to obtain a polydispersity of experimental molecule lengths. Initial conformations are typically aligned with the flow axis of the slit.  While thermally excited cavity-to-cavity jumps do take place, leading to occupancy transitions and diffusion of the polymer center of mass, this diffusion takes place over much longer time-scales than investigated here.    We explicitly exclude molecules that undergo occupancy transitions during the acquisition, and begin recording after the molecules have finished relaxing from an initially stretched state. The geometric parameters used are: 90\,nm slit height, 165\,nm cavity depth, 630\,nm cavity width, and 1.5\,$\mu$m cavity spacing.   Individual molecules are recorded for 5000 frames with a 30\,ms exposure time (150 seconds total), using a 100x oil immersion objective, a metal-halide illumination source (Xcite) and an EM-CCD  (Andor iXon).  Molecules occupying between 5 and 15 cavities were studied, while circular molecules occupied four cavities.

\section{Qualitative Observations}

\subsection{Real-time Behaviour}

     Qualitatively, we observe that ``waves'' of excess contour propagate along the molecules as they stably straddle multiple cavities, with the propagating excitations lasting several seconds.   The dynamics can be observed most conveniently in a ``kymograph" representation, showing intensity along the molecule axis versus time.   A kymograph of a typical molecule is shown in Fig.~1, as well as a display of the integrated cavity intensities over time (each coloured square represents one cavity at one frame).  Excitations, corresponding to regions of local overdensity, extend diagonally through the kymograph as the excitation propagates.  These excitations appear to emerge randomly, typically near the end of the molecule and last for several seconds.  These dynamics can also be observed in video form (see supplemental material, http://web.mit.edu/aklotz/www/SI1.gif).    We argue that these DNA waves arise due to thermal excitations that create a local contour over-density that then flows from one cavity to the next.

\subsection{Cavity Cross-Correlation Functions}

     To study quantitatively the intensity fluctuation dynamics we examine intensity cross-correlation functions between different cavities.  Using a custom Matlab program we determine the cavity center-positions from the local maxima in the projected intensity of a given single molecule recording.  The intensity is then integrated in a $3\times3$ box around the cavity center position, yielding a time-series of the intensity for each cavity over the 5000 acquired frames.  Each $N$-cavity molecule yields an $N\times5000$ matrix intensity series.   There are $N$ distinct autocorrelation functions and $N \times (N-1)/2$ unique cross-correlation functions.    We call cross-correlation functions between adjacent cavities `neighbour correlations'; cross-correlation functions between cavities one cavity apart are called `next-nearest neighbour' cross-correlation functions.

     Figure 2 shows examples of cross-correlation functions between the cavity at one end of a molecule and the subsequent four cavities.   Qualitatively, the neighbour cross-correlation is negative at short time-lag, changes sign and increases to a maximum positive correlation before decaying and reaching the noise floor at very long time lags.   At zero time lag, the correlation between two cavities is expected to be negative because if one cavity has an excess of DNA, contour conservation dictates that the other cavity is likely to have a shortage of contour. Because the excess of contour is transferred to other cavities over a finite time interval, at a later time the excess will have transferred from one cavity to the other, yielding a positive correlation in contour intensity at some small finite time. Over longer periods the correlations are expected to decay to zero. This describes the behaviour observed in Fig.~2: negative correlations at zero time that rise to some positive maximum a short time later before decaying to zero. The time lag before the positive peak generally increases as the cavities become farther apart.
     
\begin{figure}
\centering
\includegraphics[width=0.45\textwidth]{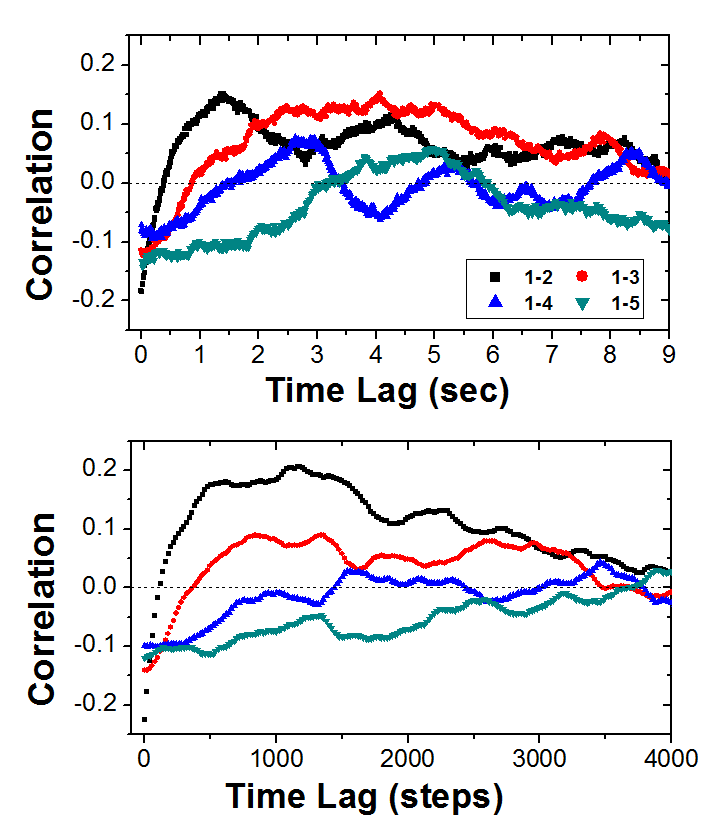}
\label{fig:sec}
\caption{Raw forward-time correlation functions between the cavity at one end of a seven-cavity spanning molecule and the second, third, fourth, and fifth cavity. The correlation with the closer cavities rises from negative to positive before decaying to an uncorrelated floor, while more distant correlations are weaker and decay more slowly.  Data is shown for an experiment (a.) and a Langevin dynamics simulation (b.)}
\end{figure}

\subsection{Cavity Autocorrelation Functions}
Measuring the autocorrelation times of the individual cavity intensities across the molecules shows that the relaxation time-scale is greater for cavities at the end of the molecule compared to cavities in the interior (see Fig. 3). The autocorrelation time was defined as the best-fit decay time of an exponential curve fit to the second-through-eleventh points of an autocorrelation function (the intercept of the fit yields the variance). To directly compare different molecules, each cavity index was normalized such that the cavities on either end had an index of $-1$ or $+1$, and the central cavity was zero (for example, in a five-cavity molecule, the two ends would have an index of $1$ and $-1$, the middle cavity would have an index of zero, and the second and fourth cavity would have an index of $\pm$0.5). By comparing the autocorrelation times of each cavity as a function of the normalized index, over an aggregate of many molecules, a trend emerges: the autocorrelation time of the end cavities is twice that of the cavities in the interior.  Averaged correlation times were equivalent to their symmetric partners on the other side of the molecule.  We also measured the single-pit autocorrelation times of circular Charomid DNA in square rings of four cavities.  Because there are no ends to break the symmetry, the averaged autocorrelation time is independent of cavity index.   The ``internal average'' of these autocorrelation times, the mean of the autocorrelation times of the non-end pit intensities, serves as a characteristic timescale of a given molecule that can be used to normalize the dynamics, compare measurements on different molecules and compare with theory.

\begin{figure}
	\centering
		\includegraphics[width=0.5\textwidth]{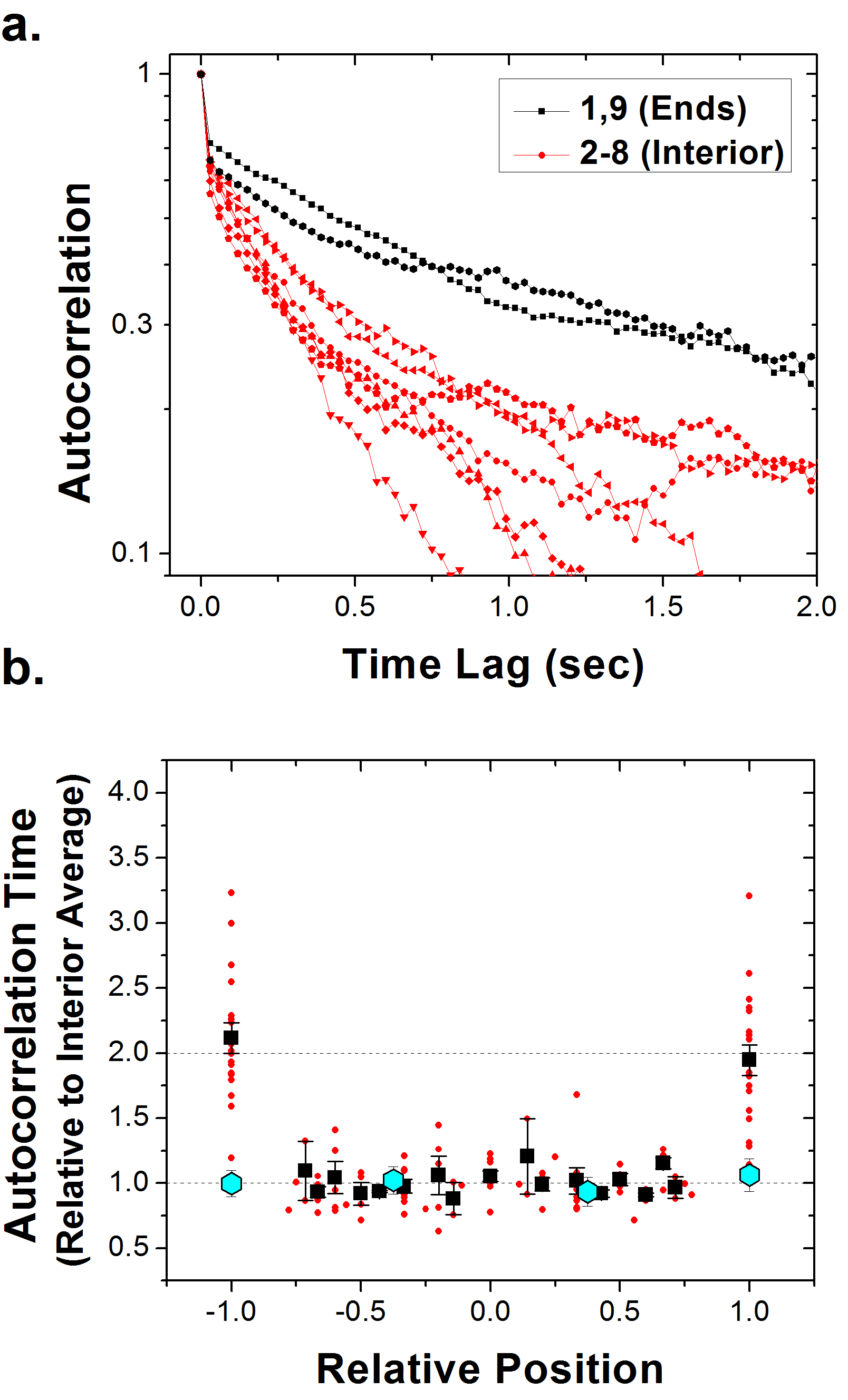}
	\label{fig:AutocorrelationTime}
	\caption{a. The autocorrelation functions of the nine in-pit intensities of a single molecule. The end-pit autocorrelation functions decay more slowly. b. The autocorrelation times of the individual cavities (red) as a function of their position in the array, normalized to the average of the interior correlation times of each molecule. The average of each position is shown (black). The average of the end-position autocorrelation times is twice that of the interior times, which are independent of position. Results for circular DNA (cyan) do not show the end-effect.}
\end{figure}

\section{Simulations}

To verify that the observed experimental phenomena were due to the underlying fluctuations in the DNA and not our experimental apparatus, Langevin dynamics simulations were used to model the system, similar to the algorithms used to understand two-pit relaxation \cite{dimers}. Using a bead-spring coarse-grained system \cite{slater2009modeling}, monomers were connected together in a linear fashion via the finitely extensive nonlinear elastic (FENE) potential \cite{kremergrest} and  a Weeks-Chandler-Anderson excluded volume \cite{weeks1978} was used to define a monomer size. Chains with lengths of 150 to 275 monomers were simulated. Bending rigidity was imposed as a three-bead harmonic potential with a spring constant chosen to yield a 5:1 persistence length to bead width ratio. The slits were defined between smooth walls separated by two monomer widths, with cavities defined as recesses in one of the walls. The parameters of the simulation are chosen not necessarily to replicate experimental conditions, but to ensure long-time occupation in a given large-N configuration. The chains were initialized spanning the pit array and allowed to relax into an occupancy state, and simulated for at least 100,000 bead-diffusion time-steps. The first 10,000 time-steps are discarded to eliminate initial-condition artifacts. The number of beads in each pit over time is analyzed the same way as the experimental intensity time series, producing analogous correlation functions.  The same qualitative features in the correlation functions are seen between the experimental data and the simulations (Fig. 2 b.).

\section{Lattice Diffusion Model}

   To develop a theoretical understanding of these fluctuations, we draw upon the tools of stochastic physics. We consider a molecule of contour length $L$ straddling $N$ cavities.  The contour length will be partitioned between the $N$ filled cavities and linker strands connecting the cavities.  In our study of a two-cavity system \cite{dimers}, we have found that the system's long-time behaviour is dominated by an `asymmetric' exchange mode whereby contour is transferred directly between the cavities, keeping the linker tension constant.  While contour can be exchanged between the cavities and the linkers via a symmetric exchange mode involving a net transfer of contour from the cavities into the slit, due to the high  stiffness of the linker, the symmetric exchange mode is suppressed relative to the asymmetric mode and occurs at much faster time-scales.  For the purposes of modelling the long time behaviour of an array of coupled cavities, we therefore assume that contour is transferred directly between nearest neighbour cavities with the contour in the linkers held fixed.  Furthermore, we argue that the time-scale of contour transfer $\tau$ should be given by the relaxation time of the asymmetric transfer mode in \cite{dimers} implying $ \tau \cong \frac{\eta}{kT}\frac{d}{w}a^{2}\ell$ \footnote{In this expression, $\eta$ is the buffer viscosity, kT is the thermal energy, d is the depth of the cavity, w is the effective molecular width, a is the cavity width, and $\ell$ is the lattice spacing}, typically 0.5 to 1 second.

      This physical picture leads directly to a lattice diffusion model \cite{chaikinlubensky} consisting of an aggregate of diffusing particles hopping stochastically between array sites. The frequency of hops from one site to the next depends on the number of hoppers in the initial site. This process can be written in the master equation formalism, describing the probability of finding a given particle at a given site at a given time. Because the probability density is directly correlated with the contour length per cavity, we can write a master equation in terms of the contour in the i$^{\mbox \scriptsize th}$ pit, $L_{i}$, and  $\gamma_{i,i+1}$ the rate of hopping from cavity $i$ to cavity $i+1$:
\begin{equation}
\frac{d L_{i}}{dt}=-(\gamma_{i,i+1}+\gamma_{i,i-1})L_{i}+\left(\gamma_{i+1,i}L_{i+1}+\gamma_{i-1,i}L_{i-1}\right).
\label{eq:master}
\end{equation}
In addition, for a linear array of cavities, the end-cavities (cavity $1$ and $N$) satisfy:
\begin{equation}
\frac{d L_1}{dt}=-\gamma_{1,2} L_1+\gamma_{2,1}L_2
\label{eq:end1}
\end{equation}
and
\begin{equation}
\frac{d L_N}{dt}=-\gamma_{N,N-1} L_N+\gamma_{N-1,N}L_{N-1}
\label{eq:end2}
\end{equation}
Detailed balance and the spatial equivalence of the lattice sites yield $\gamma_{i,i+1}=\gamma_{i+1,i}=\gamma_{i-1,i}=\gamma_{i,i-1}\equiv 1/\tau$.    Let $\delta L_i=L_i-\langle L \rangle$ where $\langle L \rangle$ is the contour stored in each cavity on average.   In addition, formally define a vector of cavity contour fluctuation values for the $N$-cavity system, $\delta \mathbf{L}=\left(\delta L_1,\cdots,\delta L_N\right)$, and let $\langle \mathbf{\delta L} | \delta \mathbf{L}_0 \rangle$ be the conditional average value of $\langle \delta \mathbf{L} \rangle$ (i.e. the average value of $\langle \delta \mathbf{L} \rangle$ given that at time $t=0$, $\langle \delta \mathbf{L} (t) \rangle=\delta \mathbf{L}_0$).  Equations~\ref{eq:master}-\ref{eq:end2} are then equivalent to the matrix system:
\begin{equation}
 \frac{d \langle \delta \mathbf{L} | \delta \mathbf{L}_0 \rangle}{dt}=\mathbf{A} \langle \delta \mathbf{L} | \delta \mathbf{L}_{0} \rangle
\label{eq:condav}
\end{equation}
with the symmetric matrices $\mathbf{A}$ given by (for a linear cavity array):
\begin{equation}
\mathbf{A}_{\mbox{\scriptsize linear}}=\gamma \left( \begin{array}{cccccc} -1 & 1 & 0 & 0 & 0& \cdots \\ 1  & -2 & 1 & 0 & 0 & \cdots \\ 0 & 1 &-2 & 1 & 0 & \cdots \\ \vdots & \vdots & \vdots & \ddots & \vdots & \vdots  \end{array} \right).
\label{eq:A}
\end{equation}
and for a circular array:
\begin{equation}
\mathbf{A}_{\mbox{\scriptsize circle}}=\gamma \left( \begin{array}{ccccccc} -2 & 1 & 0 & 0 & 0& \cdots & 1 \\ 1  & -2 & 1 & 0 & 0 & \cdots &0\\ 0 & 1 &-2 & 1 & 0 & \cdots &0 \\ \vdots & \vdots & \vdots & \ddots & \vdots & \vdots & \vdots \end{array} \right).
\label{eq:Acircle}
\end{equation}
Equation \ref{eq:condav} defines a set of linear, first order coupled ordinary differential equations.  The correlation function matrix $\langle\delta  \mathbf{L}(0) \delta \mathbf{L}(t) \rangle_{ij} \equiv \langle \delta L_i(0) \delta L_j (t) \rangle$ then has the formal solution \cite{grootmazur}:
\begin{equation}
\langle \delta \mathbf{L}(0) \delta \mathbf{L}(t) \rangle=e^{-\mathbf{A} |t|} \langle \delta \mathbf{L} \delta \mathbf{L} \rangle
\label{eq:cormatrix}
\end{equation}
with $ \langle\delta \mathbf{L}\delta \mathbf{L} \rangle_{ij} \equiv \langle \delta L_i \delta L_j \rangle$ a covariance matrix for the cavity fluctuations.  Equation~\ref{eq:A}  and Eq.~\ref{eq:cormatrix} readily lead to analytic solutions for the correlation functions of an $N$ cavity system given that the covariance matrix $\langle\delta \mathbf{L}\delta \mathbf{L} \rangle$ is known.  

   We can determine the fluctuation covariance matrix via the following argument.  Contour conservation and the requirement that the linker contour be held constant lead to:
\begin{equation}
\sum_{j=1}^N \delta L_i=0. \label{eq:Lcons}
\end{equation}
In addition, the off-diagonal terms in the covariance matrix, e.g. $\langle \delta L_i \delta L_j\rangle $ with $i \neq j$, are  \emph{equal}.   While this fact may seem surprising, given that it implies the covariance of cavities far apart are equal, realize that the covariance refers to the instantaneous correlation (correlation at zero lag time).  This instantaneous correlation cannot depend on cavity separation as our model does not include interactions that depend on cavity separation.  The cross-cavity correlation at zero lag must solely reflect contour conservation, e.g. the fact that if cavity $i$ has excess contour, the other cavities will have less, and on average--because the cavities are physically equivalent--we would expect the net loss of contour to be spread evenly across the cavities, and indeed this is observed experimentally.  Naturally, at lag times greater than zero, physical proximity of the cavities does effect the cavity correlation (with the overall details determined by the solution of Eq.~\ref{eq:cormatrix}).  Multiplying Eq.~\ref{eq:Lcons} by $\delta L_i$ averaging and solving for $\langle \delta L_i \delta L_j \rangle$ we find:
\begin{equation}
 \langle \delta L_i \delta L_j \rangle=-\frac{1}{N-1} \langle \delta L_i^2 \rangle. \label{eq:cov}
\end{equation}
Equation \ref{eq:cov} determines the elements of the covariance matrix in terms of $\langle \delta L_i^2 \rangle$, the single cavity fluctuation variance, which then determines the overall amplitude of the correlation matrix via Eq.~\ref{eq:cormatrix}.   Note that the single cavity fluctuation variance is also equal for each cavity (as the cavities have the same dimensions): for notational convenience let  $\sigma^2 \equiv \langle \delta L_i^2 \rangle$.  Equation~\ref{eq:cov} can alternatively be derived explicitly from the cavity partition function (details given in the appendix).

   Finally, note symmetries reduce the number of independent correlation functions.  Two-fold symmetry, for example, implies that correlations between symmetrically positioned pairs of cavities along the cavity array must be equal (e.g. $\left\langle \delta L_{1}(0) \cdot \delta L_{2}(t) \right\rangle$ is equivalent to $\left\langle \delta L_{N}(0) \cdot \delta L_{N-1} (t) \right\rangle$).  Microscopic reversibility implies that the autocorrelation function matrix is symmetric \cite{grootmazur}, so that  $\langle \delta L_i(0)  \cdot  \delta L_j (t) \rangle=\langle \delta L_j (0) \cdot  \delta L_i (t) \rangle$.

\section{Quantitative Results}

\subsection{End-Effects and Topology}

   As an initial test of our master equation model, we apply it to describe the observed end-effects in the autocorrelation functions (see Fig. 3).   Our model predicts that the autocorrelation functions of the end-cavities should, at short times, relax two times slower than the autocorrelation function of the interior cavities, simply because contour can escape from the end-cavities via only one adjacent cavity, while contour in the interior cavities can escape via two adjacent cavities (e.g. compare Eq.~\ref{eq:master} with Eq.~\ref{eq:end1} and Eq.~\ref{eq:end2}).  Indeed, we find experimentally that the short time relaxation time-scale of the end-cavities is twice that of the interior cavities, agreeing with the lattice model prediction.  (At longer times, our model does predict that the single cavity relaxation behaviour is more complex, with multiple exponential terms appearing in the autocorrelation functions, but we do not resolve these terms above our noise floor).  

     Circular DNA represents a unique system to explore the role of end-effects and topology in the propagation of these thermal excitations.  Because of the short length of Charomid relative to T4, we focused on states where the molecules occupied a $2\times2$ square ring of cavities.  Equation~\ref{eq:cormatrix} solved for $N=4$ yields simple solutions for the autocorrelation and the nearest neighbour cross-correlation functions:

\begin{equation}
\left\langle \delta L_{i}(0) \cdot \delta L_{i}(t) \right\rangle =\sigma^2 \left[ \frac{2}{3}e^{-2 t/\tau}+\frac{1}{3}e^{-4 t/\tau} \right]
\label{eq:circ_ac}
\end{equation}

\begin{equation}
\left\langle \delta L_{i}(0) \cdot \delta L_{i+1}(t) \right\rangle =-\frac{\sigma^2}{3}e^{-4 t/\tau}
\label{eq:circ_nn}
\end{equation}

     The predicted neighbour correlation function Eq.~\ref{eq:circ_nn} is a simple exponential decay while the autocorrelation Eq.~\ref{eq:circ_ac} is double exponential.  Examining the neighbour correlation functions of the circular molecule (see Fig. 4a), the exponential decay was  observed at short times, with a characteristic timescale of roughly 100\,ms. The zero-lag component of the exponential was found to be -0.15, roughly a factor of two smaller than the ideal -1/3 (a lower zero-time correlation arises experimentally due to noise sources such as shot noise).  Applying the decay time and zero-lag component obtained from the neighbour correlation to the double exponential autcorrelation (Eq.~\ref{eq:circ_ac}) yields a curve that follows the measured autocorrelation.  For a circular molecule, note that the autocorrelation time necessarily does not depend on cavity position (Fig.~3).  Another more subtle difference in the correlation behaviour of circular versus linear DNA for the four-cavity system is the absence of a neighbour-correlation peak for the circular molecule.  This feature is predicted in the lattice model and observed in experiment (Fig.~4b).  

\begin{figure}
	\centering
		\includegraphics[width=0.45\textwidth]{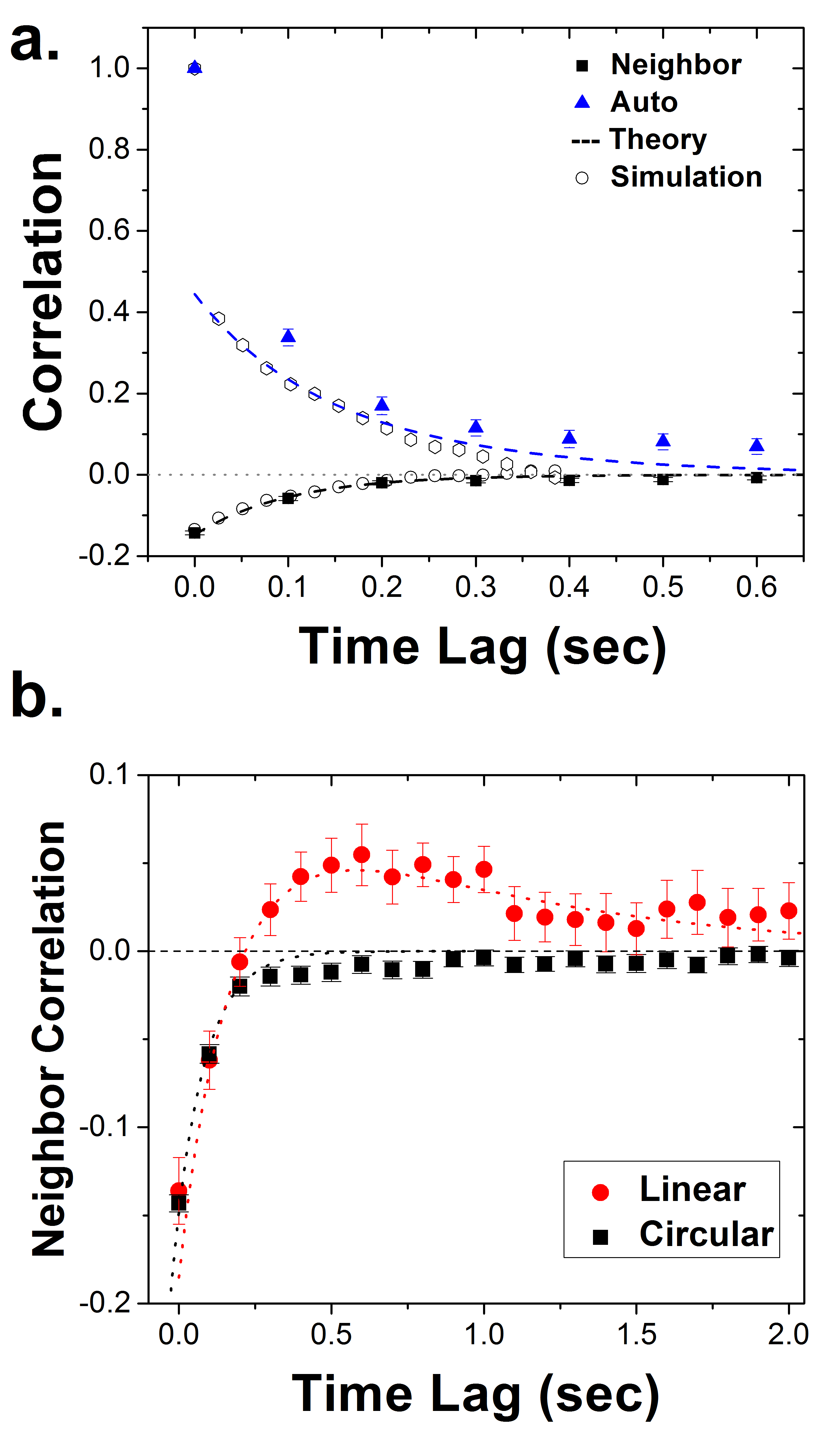}
	\label{fig:circ}
	\caption{a. The averaged autocorrelation and neighbour-correlation functions for circular DNA molecules in four-cavity rings. An exponential fit to the neighbour-correlation data is shown, and used to predict the autocorrelation function. Results from Langevin dynamics simulations are also presented. b. Comparison of the four-pit neighbour correlation functions for linear and circular DNA. There is no positive correlation peak seen in the circular data. Overlaid are the predictions of the lattice model}
\end{figure}

\subsection{Propagation Correlations}

     To examine the behaviour of excitation propagation, we focus on the specific case of molecules straddling seven cavities (see Fig. 5).  The mean of the interior autocorrelation time is used as a best estimate of the transfer time-scale $\tau$ and used to rescale the time-axis (we find $\tau=$0.47$\pm$0.02 seconds). To account for differences in the zero-lag variance between different correlation functions, we subtract the zero-lag component, so the correlation functions are normalized to zero at zero lag (we call correlation functions normalized this way ``$\Delta$-correlation functions").  A molecule spanning seven cavities has three unique correlation functions.  The ``end'' correlation functions $\left\langle L_{1}(0) \cdot L_{2}(t) \right\rangle$ and $\left\langle L_{6}(0) \cdot L_{7}(t)\right\rangle$ (1-2 and 6-7), the ``second-from-the-end'' correlation functions  $\left\langle L_{2}(0) \cdot L_{3}(t) \right\rangle$ and $\left\langle L_{5}(0)\cdot L_{6}(t) \right\rangle$ (2-3 and 5-6), and the ``middle'' correlation functions  $\left\langle L_{3}(0) \cdot L_{4}(t) \right\rangle$ and $\left\langle L_{4} (0) \cdot L_{5} (t) \right\rangle$ (3-4 and 4-5).   We average together the two equivalent cross-correlation functions measured for each molecule and then average the correlation functions again over all the observed seven-cavity molecules.
   
        Each solution of the master equation formalism for $N=7$ (Eq.~\ref{eq:cormatrix}) can be described as a sum over exponentials with coefficients corresponding to the eigenvalues of a $7\times7$ matrix.  For convenience we used the Runge-Kutta algorithm to numerically generate solutions.  With the time-axis normalized by the interior autocorrelation time, these solutions depend on only one parameter:  the single cavity variance.  We obtain the single cavity variance from exponential fits to the interior autocorrelations functions. We find $\sigma^2=0.53$, yielding good agreement between the averaged experimental data and the theoretical predictions of the lattice diffusion model (Fig. 5). 

With respect to the interior average autocorrelation time, the model predicts a correlation peak for the end correlation functions (1-2 and 6-7) at a normalized lag time of $t/\tau=2.9$, consistent with the experimental value $t/\tau=3.3\pm0.5$.  We also find good agreement between the predicted and measured correlation functions for second-from-end (2-3, 5-6) and middle (3-4, 4-5) correlation functions. The 1-2 neighbour correlation functions reach their greatest positive value at some time $\tau_{12}$, and the 2-3 neighbour correlation functions at a smaller time $\tau_{23}$. Theoretically we expect the ratio $\tau_{12}/\tau_{23}$ to be 1.4. Experimentally we observe this ratio to be 1.5$\pm$0.3.
        

\begin{figure}
	\centering
		\includegraphics[width=0.5\textwidth]{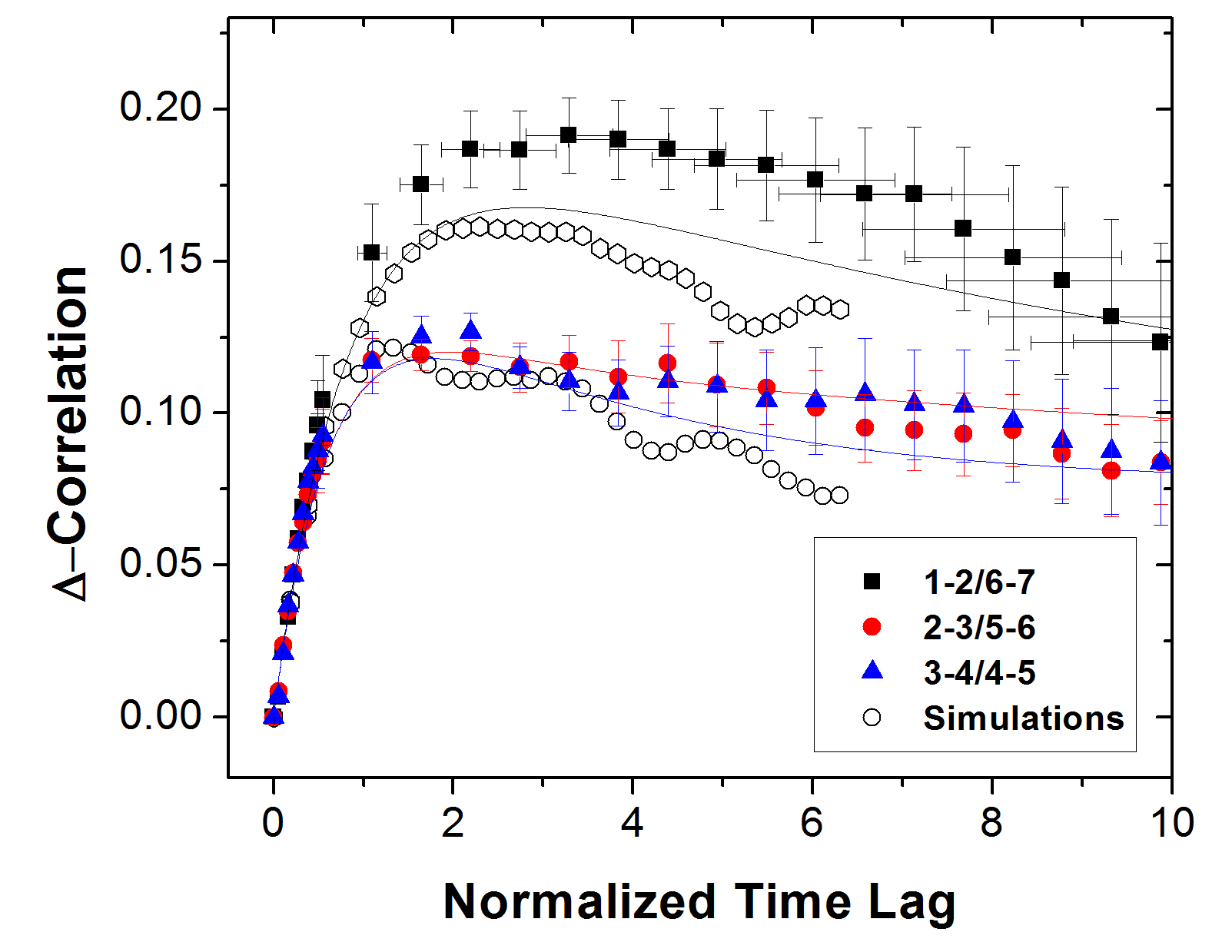}
	\label{fig:theorycompare}
	\caption{The averaged nearest neighbour $\Delta-$correlation functions of seven-cavity molecules, grouped according to their position relative to the end of the molecule.  To form the $\Delta$-correlation functions we subtract off the zero-time variance, so that the correlation at zero-lag is normalized to zero.  The time-axis is normalized by the mean autocorrelation time of the interior pits.  Overlaid are the output of the Langevin dynamics simulations (for 1-2 and 3-4) and the predictions of the lattice diffusion model (normalized in the same way).}
\end{figure}

     The autocorrelation peak provides information on how thermal contour excitations propagate across the cavity lattice.  Consider a fluctuation originating at an end-cavity.  There will be a time lag, $\tau_{12}$, corresponding to the average time for this fluctuation to propagate to the adjacent cavity; this time lag corresponds to the peak in the neighbour correlation function.  In addition, there will be a time lag, $\tau_{13}$, for the fluctuation to propagate to the next cavity over (corresponding to a peak in the next-nearest-neighbour correlation function). The ratio of the two lag time-scales, $\tau_{13}/\tau_{12}$, provides information regarding the physics of these propagations.  If they are purely diffusive then this ratio is expected to be four; if they are purely ballistic then it is expected to be two. The solutions of the master equation show that $\tau_{13}/\tau_{12}=2.64$, largely independent of $N$ if $N>5$.  We can access this ratio experimentally by determining the lag-times corresponding to the peaks of the correlation functions for all data with $N>5$ ($\left\langle L_{1}(0) \cdot L_{2}(t) \right\rangle$ and $\left\langle L_{N}(0) \cdot L_{N-1}(t)\right\rangle$ to find $\tau_{12}$ and $\left\langle L_{1}(0) \cdot L_{3}(t) \right\rangle$ and $\left\langle L_{N}(0) \cdot L_{N-2}(t) \right\rangle$ to find $\tau_{13}$).  Although there is significant variation between molecules (similar to the variation seen in figure 3 before averaging), the average value of the ratio $\tau_{13}/\tau_{12}=2.6\pm0.2$, consistent with our model prediction.  Inverting $\tau_{13}/\tau_{12}$ to find the propagated distance as a function of time yields a growth exponent close to 0.75, exactly intermediate between the diffusive and ballistic cases. The model predicts that the propagation of the excitation far from its source should be diffusive.  We do not suggest that these propagations from the end are literally ballistic, but rather that the boundary condition imposed by the linear topology causes them to propagate super-diffusively. Excitations arising in the center of the molecule do not yield a positive peak in the next-nearest neighbour correlation picture until N=9, but our larger-N data suggests that the growth exponent near the middle is 0.5$\pm$0.1, while the model predicts 0.58. The results of the ratio measurements are collated in the table below.
   \\  
      
    \begin{tabular}{|c|c|c|}
     \hline

    Ratio & Theoretical & Experimental \\
   \hline

   $ \tau_{12}/\tau_{13}|N=7$ & 1.4   & 1.5$\pm$0.3. \\
 \hline

 $ \tau_{13}/\tau_{12}|N>5$ & 2.64  & 2.6$\pm$0.2 \\
 \hline

  $ \tau_{35}/\tau_{45}|N=9$ & 0.58  & 0.5$\pm$0.1 \\
 \hline
    
    \end{tabular}%

\section{Discussion and Conclusion}

     This investigation was motivated by the observation of waves of contour propagating through long nanoconfined DNA molecules. We examined the correlation functions of the in-cavity intensities to measure the universal features of these excitations.  Langevin dynamic simulations were performed to verify that the observed fluctuations were indeed robust features of the physics and not an experimental artifact.  Finally, we interpreted the correlation functions from both experiment and simulation using a minimal model of coupled lattice diffusion that assumes a fixed time-scale for the transfer of contour from one cavity to another.    A key benefit of the nanocavity-nanoslit system is the effective compartmentalization of contour into the different cavities.  This compartmentalization of contour renormalizes the many-monomer polymer dynamics into a few-body lattice system.  Practically, rather than worrying about the global distribution of contour intensity (e.g. as was required in \cite{karp}), we can then simply consider the global intensity of each individual cavity.  The nanocavity system might serve as a model to study other collective phenomena specific to polymer physics, such as reptation, where the tendency of a large trapped molecule to undergo a transition from its ends rather than its interior (see supplemental movie 2, http://web.mit.edu/aklotz/www/SI2.gif) may effect a virtual quasi-one-dimensional tube around the molecule already in a quasi-two-dimensional slit.  Systematic measurements of the speed of fluctuation propagation under varying levels of tension may provide useful tool for studying tension propagation in confined chains.  
     
     Our model explicitly ignores a number of features, including the connectivity of the chain. Certain hopping sequences may be forbidden in the master equation if connectivity were to be respected, for example one hopper starting to the left of another and moving to the right.  When considering the Fourier modes in a nanochannel-confined DNA molecule, Karpusenko et al. \cite{karp} noted that their stochastic model broke down beyond a micron length-scale, suggesting the importance of chain connectivity at greater distances.  The fact that our connectivity-agnostic model captures the observed dynamics indicates that these subtleties are unimportant, consistent with previous findings \cite{dimers} that the time scale of tension-driven fluctuations in confined chains is much shorter than the time-scale of excluded volume-driven transfer of contour.   While it may seem curious that a model of non-interacting brownian diffusers can capture the dynamics of excluded-volume driven fluctuations, the connection can be understood by considering that hopper overpresence at a given site is Gaussianly unlikely, which is equivalent to the Boltzmann-weighted quadratic cost of excess excluded volume.  
     The explicit ignorance of tension fluctuations in the linking strand prevents the model from fully encapsulating the dynamics of these waves. For example, the potential emergence of secondary fluctuation time-scales due to a finite-time lag due to the speed of tension propagation would impart distance-dependence to the covariance matrix, and may be investigated in future experiments.  These effects might be more readily apparent in thinner nanoslits that can accommodate higher linker tension.

\textbf{Overall}, we have demonstrated that dynamic intramolecular fluctuations can be understood by considering the correlation functions that arise between different compartments of the molecule. 

\section{Appendix}

    Here we derive the covariance matrix for a molecule confined in an $N$-cavity array.  We assume that, for small fluctuations $\delta L_i$ from the equilibrium value $\langle L \rangle$ of the stored contour in each cavity, the free energy of the i$^{\mbox{\scriptsize th}}$ cavity is given by $F_i/k_B T=(1/2) A (\delta L_i)^2$ with $A$ a spring constant that depends in detail on the cavity dimensions and spacing.   Assuming that we can neglect fluctuations in the contour stored in the linker strand connecting the cavities, the spring constant is simply the spring constant of the asymmetric mode, deduced in \cite{dimers}:  $A=w/V_c$ where $w$ is the chain effective width and $V_c$ is the cavity volume $V_c=da^2$.  The partition function is then:
\begin{equation}
Z=\left[ \prod_{n=1}^N \int e^{-B (\delta L_n)^2} d(\delta L_n) \right] \delta(\sum_{i=1}^N \delta L_i )
\label{eq:Z}
\end{equation}
The covariance matrix element, for the case $i \neq j$:
\begin{eqnarray}
\langle \delta L_i \delta L_j \rangle & = &  \label{eq:Lij} \\ 
                                                       &   & \frac{1}{Z} \left[ \prod_{n=1}^N \int \delta L_i \delta L_j e^{-B (\delta L_n)^2} d(\delta L_n) \right] \delta(\sum_{i=1}^N \delta L_i ) \nonumber
\end{eqnarray}
and for $i=j$:
\begin{eqnarray}
\langle (\delta L_i)^2 \rangle & = &  \label{eq:Lii} \\ 
                                                       &   & \frac{1}{Z} \left[ \prod_{n=1}^N \int (\delta L_i)^2 e^{-B (\delta L_n)^2} d(\delta L_n) \right] \delta(\sum_{i=1}^N \delta L_i ) \nonumber
\end{eqnarray}
The delta function enforces the contour conservation constraint Eq.~\ref{eq:Lcons}.   Note that the integration limits are from $-\infty$ to $\infty$.  Equation~\ref{eq:Z}, \ref{eq:Lij} and \ref{eq:Lii} can be evaluated using the Fourier integral:
\begin{equation}
\delta(x)=\frac{1}{2 \pi} \int e^{ikx} dk. 
\label{eq:fourier}
\end{equation}
 The covariance matrix elements then follow via Gaussian integration.  For $i \neq j$,
\begin{equation}
\langle \delta L_i \delta L_j \rangle=-\frac{1}{2} \frac{1}{A N}
\end{equation}
and for $i=j$:
\begin{equation}
\langle (\delta L_i)^2 \rangle=\frac{1}{2} \frac{N-1}{A N}.
\end{equation}
These relations are consistent with Eq.~\ref{eq:cov}.

\section{Acknowledgements}
The authors wish to thank Lyndon Duong and Mikhail Mamaev for help with experiments. This work was supported by a National Science and Engineering Research Council of Canada Discovery Grant (NSERC-DG, 386212-10) with equipment provided by an NSERC tools grant and the Canada Foundation for Innovation (CFI).  In addition, the authors thank Laboratoire de Micro- et Nanofabrication (LMN) at INRS-Varennes and McGill Nanotools-Microfab for supporting the nanofabrication.  We thank the facility for electron microscope research at McGill (FEMR) for providing access to the scanning electron microscope. HdH gratefully acknowledges funding from NSERC via a Discovery Grant (2014-06091). ARK is supported by an NSERC postdoctoral fellowship.

\end{document}